\documentstyle{l-aa}
\begin{document}
\thesaurus{08(08.22.3,08.06.3)}
\title{RR Lyrae Variables in the Globular Cluster M5}
\author{J. Kaluzny\inst{1}, ~A. Olech\inst{2}, ~I. Thompson\inst{3}, 
~W. Pych\inst{2},
~W. Krzeminski\inst{1,3} \and ~A. Schwarzenberg-Czerny\inst{1,4}}
\offprints{Arkadiusz Olech, e-mail: olech@sirius.astrouw.edu.pl}
\institute{
Copernicus Astronomical Center,
ul. Bartycka 18, 00-716 Warsaw, Poland (jka,wk,alex@camk.edu.pl)
\and
Warsaw University Observatory, Al. Ujazdowskie 4, 00-478
Warsaw, Poland (olech,pych@sirius.astrouw.edu.pl)
\and
Carnegie Institution of Washington, 813 Santa Barbara Street, Pasadena,
CA 91101, USA (ian@ociw.edu)
\and
Astronomical Observatory of Adam Mickiewicz University,
ul. Sloneczna 36, 61-286 Poznan, Poland}
 
\date{Received ................., 1999, Accepted ..............., 1999}
 
\maketitle
\markboth{Kaluzny {\it et al. } RR Lyr Variables in M5}{}
\begin{abstract}

We present $V$-band CCD photometry of 65 RR Lyr variables from the
globular cluster M5. We have estimated the basic physical parameters
for 16 RRc stars and 26 RRab stars using a Fourier decomposition of the
light curves of the variables. The mean values of mass, luminosity,
effective temperature and relative helium abundance for the RRc stars
are measured to be ${\cal M}=0.54~{\cal M}_\odot$, $\log
(L/L_\odot)=1.69$, $T_{eff}=7353$~K and $Y=0.28$, respectively. For the
RRab variables the derived mean values of absolute magnitude,
metallicity and effective temperature are: $M_V=0.81$, ${\rm
[Fe/H]}=-1.23$ and $T_{eff}=6465$~K. We find that the $V$ amplitude of
an RRab star for a given period is a function of metal abundance rather
than Oosterhoff type.

We find significant problems with the calibration of both the zero point
and the scale of the luminosities measured with the Fourier technique.
The apparent distance modulus derived from RRc stars is equal to
$14.47\pm0.11$ and it is in good agreement with recent determinations.
On the other hand distance modulus obtained from the sample of RRab
stars (calibrated by the Baade-Wesselink observations of field RR Lyr
variables) is significantly smaller and equal to $14.27\pm0.04$

\end{abstract}

\section{Introduction}

NGC~5904=M5=C1516+022 is one of the richest globular clusters in the
northern hemisphere. It is placed only 7 kpc from the Sun, 5.5 kpc from
the Galactic Center and 4.9 kpc above the plane of the Galactic disc
(Zinn 1985), and thus has low reddening, $E(B-V)=0.03$ (Harris 1996).
The above properties make M5 an excellent target for detailed studies,
especially in the field of variable stars.

Spectroscopic investigations of M5 yield metallicity determinations
varying from ${\rm [Fe/H]}=-1.0$ (Butler 1975) to ${\rm [Fe/H]}=-1.4$
(Zinn 1985). Carretta and Gratton (1997) obtained ${\rm
[Fe/H]}=-1.11\pm 0.03$ on their newly introduced scale of GC
metallicities.

M5 is known to contain more than 140 RR Lyr variables (Clement 1997,
Sandquist et al.  1996, Caputo et al.  1999). CCD $VI$ photometry for 49
of them was recently presented by Reid (1996) who gives  references to
earlier papers dealing with RR Lyr stars in M5.  Recently,  Kaluzny et
al. (1999) reported the discovery of 5 faint  variables in the cluster
field: four SX Phe stars and one eclipsing binary. They also showed
that several candidate eclipsing binaries identified by Reid (1996) and
Yan and Reid (1996) are non-variable objects.   Drissen and Shara
(1998)  used $HST$ images to look for variable stars in the core of M5.
They identified one variable blue straggler, probably an eclipsing
binary,  and several new RR Lyr stars. New $BV$ photometry of the M5 RR
Lyr variables was also reported by Caputo et al. (1999). And recently
Olech et al (1999b) using newly developed image subtraction method
discovered four new variables in this cluster.

In this paper we present results based on time-series photometry
of RR Lyr stars in M5 obtained in 1997 at Las Campanas Observatory. The
detailed description of the observational runs and reduction process
were published elsewhere (Kaluzny et al. 1999). In short, the observing
material consisted of 275 $V$-band frames collected during the period
from May 9 through August 13 1997. Most of the data were taken during a
single sub-run  in May 1997. Two fields with a significant overlap
at the cluster center were monitored.

\section{Results}

Our search for variable stars in M5 has identified 65 RR Lyrae
variables. All these stars were previously known (Sawyer Hogg 1973,
Sandquist et al. 1996, Clement 1997). We have detected 49 fundamental
mode pulsators (Bailey type RRab), 15 first overtone pulsators (Bailey
type RRc), and one possible second mode pulsator (Bailey type RRe).
In this paper we use names assigned by Sawyer Hogg (1973) with the
exception of V963 (Zhukov 1971).

We fitted our $V$-band light curves with Fourier series of the form:
 
\begin{equation}
V = A_0 + \sum^{8}_{j=1} A_j\cdot\sin(j\omega t + \phi_j)
\end{equation}
 
\noindent where $\omega=2\pi/P$ and $P$ is the pulsation period of the
star. Least squares fits were computed by orthogonal projections  onto
trigonometric polynomials using a method developed by
Schwarzenberg-Czerny (1997) and Schwarzenberg-Czerny and Kaluzny (1998).
In this way we determined the values of $\omega$, $A_{j}$ and $\phi_{j}$
together with their errors, and we calculated  amplitude
combinations $R_{j1}=R_j/R_1$ and phase differences defined as
$\phi_{j1}=\phi_j-j\cdot\phi_1$. These quantities will be used in the 
following analysis.

Using the derived periods we constructed
phased $V$-band light curves  which are presented in Fig. 1. The periods
of the cluster RRc stars range from  0.2648 to 0.4325  days with a mean
period of 0.3250 days. The periods of the RRab variables are between 0.4497
and 0.8453 days with a mean value of 0.554 days. These properties place
M5 among the Oosterhoff type I clusters.

In Fig. 2 we compare our photometry with observations given by  Reid
(1996). We plot $A_0$ -- $<V>$(Reid) $vs$ $A_0$, where $A_0$ is the
mean magnitude as measured by the Fourier decomposition of the light
curves. Some stars in common to the two samples have not been plotted
(V6, V54, and V91:  our light curve or Reid's measured with
particularly poor signal-to-noise; V24 and V52: Reid's data do not
sample the maxima of the light curves). For the remaining 32
measurements in common the mean value of $A_0$ -- $<V>$(Reid) = --0.011
$\pm$ 0.026.

In Fig. 3 we present  the relations between the pulsational period  and
the peak to peak $A_V$ amplitude, $R_{21}$, $\phi_{21}$, $\phi_{31}$
and $\phi_{41}$. In these plots RRc stars occupy locations quite
distinct from RRab stars, and in the following sections we discuss
these two groups of stars separately. The relative lack of scatter in
the relations presented in Fig. 3 attests to the overall quality of our
light curves (cf. Clement et al. 1992, Simon and Clement 1993).

\begin{table*}
\caption[ ]{Light curve parameters for RRc variables from M5}
\begin{flushleft}
\begin{tabular}{|r|c|c|c|c|c|c|c|c|c|c|c|c|}
\hline
\hline
Star & Period & $A_V$ & $A_0$ & $A_1$ & $R_{21}$ & $\sigma_{R_{21}}$ &
$\phi_{21}$ & $\sigma_{\phi_{21}}$ & $\phi_{31}$ & $\sigma_{\phi_{31}}$ &
$\phi_{41}$ & $\sigma_{\phi_{41}}$ \\
\hline
\hline
V15  & 0.336776 & 0.41 & 15.064 & 0.206 & 0.087 & 0.010 & 3.067 & 0.117  &  7.130 & 0.142  & 3.571 & 0.243\\
V31  & 0.300575 & 0.50 & 15.069 & 0.261 & 0.157 & 0.004 & 3.100 & 0.019  &  6.117 & 0.028  & 3.356 & 0.042\\
V35  & 0.308177 & 0.46 & 14.967 & 0.234 & 0.154 & 0.004 & 3.431 & 0.045  &  6.542 & 0.073  & 3.644 & 0.106\\
V40  & 0.317334 & 0.43 & 15.051 & 0.220 & 0.114 & 0.005 & 3.147 & 0.034  &  6.549 & 0.046  & 3.975 & 0.091\\
V44  & 0.329576 & 0.42 & 15.027 & 0.214 & 0.107 & 0.009 & 2.899 & 0.080  &  6.967 & 0.128  & 4.026 & 0.259\\
V55  & 0.328938 & 0.41 & 15.086 & 0.210 & 0.119 & 0.010 & 3.438 & 0.080  &  7.132 & 0.133  & 4.024 & 0.188\\
V57  & 0.284673 & 0.49 & 15.050 & 0.246 & 0.179 & 0.004 & 3.305 & 0.032  &  6.065 & 0.065  & 3.256 & 0.076\\
V62  & 0.281412 & 0.53 & 15.099 & 0.262 & 0.187 & 0.004 & 3.107 & 0.030  &  5.808 & 0.067  & 3.036 & 0.079\\
V66  & 0.350658 & 0.44 & 15.032 & 0.225 & 0.053 & 0.005 & 3.299 & 0.122  &  7.343 & 0.095  & 4.113 & 0.216\\
V73  & 0.340111 & 0.51 & 14.974 & 0.262 & 0.095 & 0.008 & 3.364 & 0.085  &  7.047 & 0.098  & 3.854 & 0.150\\
V76  & 0.432544 & 0.40 & 14.828 & 0.192 & 0.036 & 0.005 & 5.524 & 0.201  &  8.276 & 0.109  & 5.315 & 0.212\\
V78  & 0.264798 & 0.39 & 15.094 & 0.196 & 0.158 & 0.005 & 3.090 & 0.038  &  6.209 & 0.137  & 2.698 & 0.173\\
V79  & 0.333089 & 0.35 & 14.947 & 0.177 & 0.147 & 0.011 & 3.134 & 0.151  &  7.033 & 0.237  & 3.379 & 0.388\\
V80  & 0.336549 & 0.39 & 15.042 & 0.196 & 0.117 & 0.005 & 2.998 & 0.057  &  7.194 & 0.116  & 4.014 & 0.146\\
V88  & 0.328070 & 0.43 & 15.034 & 0.220 & 0.123 & 0.014 & 3.496 & 0.100  &  6.626 & 0.183  & 4.045 & 0.285\\
V130 & 0.326624 & 0.42 & 14.882 & 0.212 & 0.033 & 0.028 & 4.460 & 0.876  &  7.544 & 0.395  & 4.464 & 0.776\\
\hline
Mean & 0.324994 & 0.44 & 15.015 & 0.221 & 0.117 & -     & 3.429 & -      &  6.849 & -      & 3.798 & -    \\
\hline
\hline
\end{tabular}
\end{flushleft}
\end{table*}

\begin{table}
\caption[ ]{Physical parameters derived for RRc Lyrae variables from M5}
\begin{flushleft}
\begin{tabular}{|r|c|c|c|c|c|r|c|}
\hline
\hline
Star & Mass & $\sigma_{\rm Mass}$ & $\log L$ & $\sigma_{\log L}$ & $T_{eff}$ &
$\sigma_{T}$ & $Y$\\
\hline
\hline
V15  & 0.508 & 0.018  & 1.687 & 0.008 & 7338 &13 & 0.285\\
V31  & 0.618 & 0.004  & 1.694 & 0.002 & 7363 & 2 & 0.276\\
V35  & 0.563 & 0.010  & 1.681 & 0.004 & 7377 & 6 & 0.283\\
V40  & 0.570 & 0.007  & 1.694 & 0.003 & 7346 & 4 & 0.279\\
V44  & 0.523 & 0.017  & 1.687 & 0.007 & 7345 &12 & 0.284\\
V55  & 0.501 & 0.017  & 1.676 & 0.008 & 7363 &12 & 0.288\\
V57  & 0.609 & 0.010  & 1.673 & 0.004 & 7416 & 6 & 0.283\\
V62  & 0.646 & 0.011  & 1.683 & 0.004 & 7404 & 6 & 0.278\\
V66  & 0.491 & 0.012  & 1.693 & 0.006 & 7315 & 8 & 0.284\\
V73  & 0.521 & 0.013  & 1.696 & 0.006 & 7320 & 9 & 0.281\\
V76  & 0.432 & 0.012  & 1.734 & 0.006 & 7182 &10 & 0.277\\
V78  & 0.566 & 0.020  & 1.632 & 0.008 & 7509 &13 & 0.297\\
V79  & 0.517 & 0.031  & 1.688 & 0.014 & 7340 &22 & 0.284\\
V80  & 0.499 & 0.015  & 1.683 & 0.007 & 7345 &10 & 0.286\\
V88  & 0.569 & 0.026  & 1.704 & 0.011 & 7318 &17 & 0.276\\
V130 & 0.450 & 0.045  & 1.649 & 0.023 & 7410 &37 & 0.300\\
\hline
\hline
\end{tabular}
\end{flushleft}
\end{table}

\begin{table*}
\caption[ ]{Mean parameters for RRc stars in
globular clusters after Clement and
Shelton (1997) and Kaluzny et al.  (1998).}
\begin{flushleft}
\begin{tabular}{|l|c|c|c|c|c|c|c|}
\hline
\hline
Cluster & Oosterhoff & [Fe/H] & No. of & mean & mean & mean & mean \\
        &   type     &        & stars  & mass & $\log L$ & $T_{eff}$ &$Y$ \\
\hline
\hline
NGC 6171 & I & -0.68 & 6 & 0.53 & 1.65 & 7447 & 0.29\\
M5       & I & -1.25 & 7 & 0.58 & 1.68 & 7388 & 0.28\\
M5       & I & -1.25 &14 & $0.54\pm0.02$ & $1.69\pm0.01$ & $7353\pm19$ & $0.28\pm0.01$ \\      
M3       & I & -1.47 & 5 & 0.59 & 1.71 & 7315 & 0.27\\
M9       & II& -1.72 & 1 & 0.60 & 1.72 & 7299 & 0.27\\
M55      & II& -1.90 & 5 & 0.53 & 1.75 & 7193 & 0.27\\
NGC 2298 & II& -1.90 & 2 & 0.59 & 1.75 & 7200 & 0.26\\
M68      & II& -2.03 &16 & 0.70 & 1.79 & 7145 & 0.25\\
M15      & II& -2.28 & 6 & 0.73 & 1.80 & 7136 & 0.25\\
\hline
\hline
\end{tabular}
\end{flushleft}
\end{table*}

\subsection{RRc stars}

The resulting Fourier parameters for RRc stars are presented in Table 1.
In the following sections we discuss individual RRc stars with atypical
light curves, and then the derivation of physical parameters of the 
RRc stars from the Fourier coefficients.

\subsubsection {Individual stars}

The light curve of V130 reveals scatter which is about 10 times larger
than the scatter observed for most variables. However, prewhitening 
the observations
with the base period and its 3 harmonics leaves no power in excess of 0.02
mag amplitude in the frequency range  $0$ -- $300 c/d$. Inspection of
the frames reveals 3 nearby companion stars, and we conclude that
the scatter caused by crowding. The shift in the light curve of
V88 also appears to be instrumental in origin (our light curves were
obtained from two overlapping fields and some stars laying at large
distance from the overlapping region may have small shift between light
curves obtained from different fields). This star has a frequency $3.05
c/d$, so that the phasing of nearby observations is similar. Removal of
the base frequency and its harmonics leaves a very low frequency
residual signal.

Two stars are outliers in the $A_V$--$\log P$, $R_{21}$--$\log P$,
$\phi_{21}$--$\log P$ and $\phi_{31}$--$\log P$ relations presented in
Fig. 3.  One of these is V76 --  the RRc variable with the longest
period in our sample at 0.4325 d. 
Another outlying object is V78 -- the star with the shortest
period in our sample.  Recently Minniti et al.  (1997) published a
study of the  RR Lyr variables  in the  MACHO Collaboration database.
They found three peaks in the period distribution of stars in the Large
Magellanic Cloud and in the Galactic bulge. The two most prominent
peaks are the RRab and RRc pulsators and the lowest peak, at a period
of about 0.27-0.28 $d$, was interpreted as  due to RRe stars
(pulsations in the 2nd overtone). A similar result for RR Lyr variables
from the Galactic bulge was obtained by Olech (1997) and for variables
from the globular cluster IC 4499 by Walker and Nemec (1996).
Theoretical calculations performed by Sandage (1981) support the
hypothesis that RR Lyr stars with the shortest periods and smallest
amplitudes may be RRe type pulsators. Additionally Stellingwerf et al.
(1987) predicted that, if RRe stars exist, they should have light
curves that have a sharper peak at maximum light than the 1st overtone
pulsators. Variable V78 fits this description very well. It has the
shortest period among our sample of RR Lyr stars, its amplitude is 
low and its light curve (see Fig. 1) is more asymmetric than the light
curves of other RRc stars.

The light curves of other RRc variables with the shortest periods, from
$0.28^d$ to $0.30^d$, i.e. V31, V35, V57 and V62, are  remarkably
similar both in their general shape and in fine details. It is perhaps
interesting that they all display small bumps just before the maxima
and that the scatter of observations around the maxima appears to be
slightly larger than the scatter at minimum light.
Light curves of other stars with periods just
above the range 0.28-0.30$^d$ do not show such behavior.  Because of
this consistency, instrumental origin is  not a likely cause of these
effects (e.g. saturation of the CCD images at maximum light). If
confirmed by further observations of comparable accuracy, these effects
could reveal potentially interesting dynamical processes.  We have
already argued that V78, the star with the shortest period in our
sample, is likely to pulsate in the second overtone.  Hence, one has to
consider a possibility that some sort of interaction between the first
and second overtones is responsible for the effects observed in the
light curves of  V31, V35, V57 and V62.

\subsubsection{Physical Parameters}

Simon and Teays (1982), Simon (1989) and Simon and Clement (1993)
(hereafter SC) have presented a method of estimating the masses,
luminosities, effective temperatures and helium abundances of RRc stars
based only on the Fourier parameters of  $V$-band light curves. The
relevant equations are summarized in Olech et al. (1999a). The existence
of a relation between the masses, luminosities, temperatures  and
metallicities of pulsating stars and their Fourier parameters is based
on hydrodynamic pulsation models. However, the details of the specific
calibrations based on models are  still subject to revisions.  Hence, we
list in Table 1 the cluster averages and standard deviations of the Fourier
parameters as model-free characteristics of the pulsating stars in this
particular cluster. Please note that the standard deviations refer to
the spread of derived  parameters and not to observational uncertainties 
(errors of the Fourier  parameters are listed in Table 1).

Table 2 presents the estimated  masses, luminosities, effective
temperatures and helium abundances for all of the RRc stars in our
sample.  The errors presented in Table 2 are calculated from the error
propagation law. Only two stars from Table 2 have errors of $\phi_{31}$
larger than 0.2 (V79 and V130), and we have omitted them in further
analysis.

In Fig. 4 we have plotted the calculated values of $\log (L/L_\odot)$
(hereinafter $\log L$)  against the observed values of $A_0$ (i.e. the
mean observed magnitude). The solid lines have a slope of 0.4 and are
separated by 0.04 in $\log L$, which represents the one sigma standard
deviation in the computed values of luminosity (Simon and Clement
1993). Simon and Clement suggested that some of the scatter in this
plot for other clusters might be explained by poor photometry resulting
from crowding effects in the dense, central regions of the clusters.  We
have tested this for our sample by plotting in Fig. 4 symbols with size
proportional to the distance of a star from the center of the cluster.
It can be seen that the most discrepant point is from the center of the
cluster (this point is V130 which, as mentioned in the previous
section, is badly crowded). The remaining points lay near the solid
lines buy not between them which may suggests that the problems with
crowding do not explain fully the scatter in the $\log L-A_0$ plot. The
similar situation is in M55 where Olech et al (1999a) found that the
scatter in the $\log L vs. A_0$ relation was not due to crowding
effects.

Table 3 presents the mean parameters for RRc stars from several
clusters. This table is taken from Kaluzny et al. (1998). A previous
determination of the physical parameters of RRc stars in M5 was made by
Clement and Shelton (1997) who used the data of Reid (1996). Only seven
stars from Reid's sample had errors of $\phi_{31}$ smaller than 0.2.
Our sample is twice as large, and therefore the derived mean values of
mass, luminosity, temperature and helium abundance are statistically
better defined in comparison with the results of Clement and Shelton
(1997).
 
Clement and Shelton (1997) noted the existence of correlations between
the mean values of several parameters for RRc stars belonging to
different clusters. Specifically, an increasing mean mass corresponds
to an increasing value of luminosity and a decreasing value of
effective temperature and helium abundance.  Our results for M5 are
consistent with these correlations.  On the other hand Table 3 contains
the recent result of Olech et al. (1999a) who found that the RRc variables
in M55 have a mean mass too small to fit well into the sequence in
Table 3.

\subsection{RRab stars}

Several RRab stars exhibit quite a large scatter in their phased light
curves. In the case of V63, this is is likely due to
poor phase coverage and phasing uncertainty. V2, V4, V8, V14, V27, V65
and V89 display modulation of their light curves which reflects
intrinsic variability. The variable V91 is placed in the vicinity of a
saturated star and thus its photometry is of poor quality.

The morphology of the light curves is seen to change with period. Two
RRab stars with the shortest periods, V4 and V29, reveal broad flat
minima and fairly symmetric triangular maxima. Stars with periods of
intermediate length have a much steeper rise than decline and their minima
are often deformed by bumps. Among long period RRab stars V963 with a
small amplitude is rather special. Light curves of other long period 
stars V43, V75, V9, V87 and V77 all reveal  a break in steepness of
their rise to the maximum.

Kovacs and Jurcsik (1996, 1997, and references quoted therein,
hereafter KJ) have extended the Fourier analysis of Simon, Teays and
Clement to RRab stars.  Although their method is still  evolving, to
retain comparability with earlier results we have used their original
formulae as listed by Olech et al. (1999a).  Table 4 gives the light
curve parameters  of M5 RRab variables in our sample obtained from
Fourier fitting. As for the RRc stars, we also list cluster
averages and standard deviations of the Fourier parameters as
model-free characteristics of the whole cluster.

We have applied the KJ formulae to the values from Table 4. The results
are listed in Table 5, which contains values and errors of the absolute
magnitude $M_V$, metallicity [Fe/H], effective temperature $T_{eff}$.
We also list the deviation parameter $D_m$ as calculated from new
equations given by Kovacs and Kanbur (1998). This parameter measures
the regularity of the light curve, and according to the original paper
of KJ, their equations are valid only for RR Lyr stars with $D_m<3$. A
total of  26 stars from our sample satisfy this condition. In the
$A_V$--$\log P$ plot presented in Fig. 3 the RRab variables with
$D_m<3$ are plotted with solid triangles and these with $D_m>3$ with
open triangles. The solid line in this plot represents a linear fit to
RRab variables in M3 (Kaluzny et al. 1998).

Recently Clement and Shelton (1999) examined the $A_V$--$\log P$
relations for RRab stars in the Oosterhoff type I clusters M3 and M107
and Oosterhoff type II clusters M9 and M68. They suggested that the $V$
amplitude for a given period is not a function of metal abundance but
rather a function of the Oosterhoff type. For example, the clusters M3
and M5, which have the same Oosterhoff type, should have identical
$A_V$--$\log P$ relations.  Inspection of the upper panel of Fig. 3
shows that this is not true. A linear fit to RRab variables in M3
presented in this figure as a solid line clearly divides the regular
RRab stars into two groups. The mean magnitude of the six variables
laying above the M3 fiducial line is $15.00\pm0.02$ and the mean
magnitude of the regular RRab stars laying under the solid line is
$15.10\pm0.01$. This strongly suggests that these six RRab stars are in
a more advanced evolutionary state than the others. Similar behavior
was seen by Kaluzny et al. (1998) among the RRab stars in M3. After
excluding these six stars from our analysis, one can still see that the
remaining RRab stars do not fit the relation derived for RRab variables
from M3 but lie below it (toward lower amplitudes and shorter periods).
This suggests that the zero point of the linear $A_V$--$\log P$
relation depends on the metallicity of the cluster and not the
Oosterhoff type as suggested by Clement and Shelton (1999).

For 26 RRab stars with $D_{m}<3$ we obtained the following mean
parameters: $M_V=0.81$, ${\rm [Fe/H]}=-1.23$ and $T_{eff}=6465$~K.
These values for M5 fit well into the sequence of physical
parameters measured for other clusters with Fourier analysis as 
summarized in Table 6.

In Fig. 5 we plot $M_V$ values calculated for all our RRab stars
against  $A_0$. Again the RRab variables with $D_m < 3$ are plotted
with solid triangles and those with $D_m > 3$ with open triangles. The
solid lines with a slope of unity are separated by 0.1 mag, and
represent the one sigma uncertainty in the estimation of $M_V$ using
the KJ formalism. There are three stars (V4, V13, V54) among the RRab
variables with regular light curves (i.e. with $D_m < 3$) which are
about 0.05 mag too bright for their absolute magnitude and do not lay
between the envelope lines in Fig. 5.  These objects are also the stars
with $D_m<3$ that are closest to the center of the cluster and their
true visual magnitudes may be slightly fainter than the values derived
from our data.

The range of temperature for the RRc and RRab stars is
consistent with that presented by Reid (1996) who calculated
temperatures both from pulsation theory (Carney et al. 1992) and from
$(V-I)$ colors.

\subsection {The Distance Modulus to M5}

In Fig. 6 we plot the apparent distance moduli of the RRab and RRc stars
in our sample as measured by the Fourier technique. For the RRab stars
$M_V$ is taken from Table 5. For the RRc stars, $M_V$ has been
calculated from $\log L$ (Table 2), with $M_V$ = $M_{\rm BOL}$ -- BC, and
taking BC = 0.06[Fe/H] + 0.06 (Sandage and Cacciari 1990). We adopt 
[Fe/H] = -1.23 (see Table 6).

The different zero point calibrations of the RRc stars by SC and the
RRab stars by JK is clear from Figure 6. The zero points of the RRc
stars are calibrated directly from the hydrodynamical models, while the
RRab stars are calibrated by Baade-Wesselink observations of field RR
Lyr stars. The mean apparent distance modulus as measured from the RRc
stars (dropping V76 and V78, see discussion in Section 2.1.1) is  $(m
-M)_V$ = 14.47 $\pm$ 0.11. For the RRab stars with $D_m < 3.0$ the mean
apparent distance modulus is $(m - M)_V$ = 14.27 $\pm$ 0.04. While the
discrepancy arising from the zero point calibration of the Fourier
technique is substantial, the distance modulus obtained for RRab stars
is consistent with the value of $(m-M)_V$ = 14.33 (Harris 1996). On the
other hand recent determinations of distance modulus to M5 give the
values between 14.41 and 14.58 (Sandquist et al, 1996, Reid, 1998). The
distance modulus derived by us from RRc stars agrees within the errors
with these recent determinations. It indicates that hydrodynamical
models of Simon and Clement (1993) give the correct luminosities of RRc
stars.

It is interesting to note that the apparent distance moduli for
individual stars in Fig. 6 appears to be correlated with $A_0$ for both
the RRab and RRc star samples. Scale errors in $A_0$ are not likely to
be the explanation since the range in $A_0$ is small. In addition,
while there does seem to be a similar correlation shown in the
comparison with Reid's (1996) photometry for the small sample of RRc
stars in common between the two studies (see Fig. 2), no such
correlation is seen for the RRab stars.  The slope of these
correlations is substantial, and could compromise attempts to use the
Fourier analysis of RR Lyr light curves to derive distances to globular
clusters. Accurate CCD photometry of samples of cluster RR Lyrae stars
is needed to address this issue.

\begin{table*}
\caption[ ]{Light curve parameters for the RRab Lyrae variables in M5}
\begin{flushleft}
\begin{tabular}{|r|c|c|c|c|c|c|c|c|c|c|c|c|}
\hline
\hline
Star & Period & $A_V$ & $A_0$ & $A_1$ & $R_{21}$ &
$\sigma_{R_{21}}$ &
$\phi_{21}$ & $\sigma_{\phi_{21}}$ & $\phi_{31}$ & $\sigma_{\phi_{31}}$
& $\phi_{41}$ & $\sigma_{\phi_{41}}$ \\
\hline
\hline
V1  & 0.521777 & 1.11 & 15.145 & 0.366 & 0.500 & 0.003  & 2.351 & 0.033  & 5.026 & 0.049  & 1.394 & 0.067\\
V2  & 0.526612 & 1.09 & 15.141 & 0.393 & 0.514 & 0.009  & 2.405 & 0.084  & 5.049 & 0.125  & 1.394 & 0.165\\
V3  & 0.600149 & 0.71 & 15.076 & 0.254 & 0.480 & 0.004  & 2.463 & 0.019  & 5.311 & 0.030  & 2.007 & 0.044\\
V4  & 0.449699 & 0.96 & 15.099 & 0.395 & 0.425 & 0.008  & 2.356 & 0.100  & 4.786 & 0.152  & 1.239 & 0.216\\
V5  & 0.545824 & 1.06 & 15.099 & 0.358 & 0.447 & 0.024  & 2.157 & 0.040  & 4.976 & 0.065  & 1.555 & 0.107\\
V6  & 0.548891 & 0.92 & 15.138 & 0.301 & 0.468 & 0.033  & 2.397 & 0.109  & 5.373 & 0.159  & 1.748 & 0.241\\
V8  & 0.546143 & 0.96 & 15.120 & 0.340 & 0.462 & 0.007  & 2.350 & 0.028  & 5.037 & 0.034  & 1.513 & 0.051\\
V9  & 0.698907 & 0.80 & 14.931 & 0.284 & 0.496 & 0.004  & 2.756 & 0.031  & 5.676 & 0.048  & 2.319 & 0.073\\
V10 & 0.530662 & 1.10 & 15.133 & 0.365 & 0.490 & 0.006  & 2.290 & 0.037  & 4.952 & 0.057  & 1.338 & 0.079\\
V11 & 0.595911 & 1.11 & 14.994 & 0.380 & 0.524 & 0.003  & 2.456 & 0.008  & 5.188 & 0.013  & 1.742 & 0.017\\
V12 & 0.467707 & 1.31 & 15.165 & 0.438 & 0.463 & 0.005  & 2.221 & 0.021  & 4.760 & 0.030  & 1.141 & 0.043\\
V13 & 0.513002 & 1.11 & 14.971 & 0.376 & 0.463 & 0.014  & 2.148 & 0.051  & 4.700 & 0.071  & 1.153 & 0.103\\
V14 & 0.487172 & 1.05 & 15.130 & 0.347 & 0.487 & 0.016  & 2.124 & 0.112  & 4.726 & 0.163  & 1.061 & 0.231\\
V16 & 0.647634 & 1.19 & 14.863 & 0.399 & 0.551 & 0.006  & 2.570 & 0.014  & 5.230 & 0.021  & 1.999 & 0.030\\
V18 & 0.463956 & 1.23 & 15.151 & 0.484 & 0.380 & 0.007  & 2.191 & 0.024  & 4.527 & 0.035  & 0.771 & 0.052\\
V19 & 0.469918 & 1.34 & 15.173 & 0.484 & 0.413 & 0.007  & 2.347 & 0.043  & 4.713 & 0.060  & 1.026 & 0.083\\
V20 & 0.609551 & 0.91 & 15.058 & 0.320 & 0.509 & 0.008  & 2.493 & 0.032  & 5.253 & 0.050  & 1.998 & 0.070\\
V21 & 0.604896 & 0.98 & 15.045 & 0.330 & 0.539 & 0.003  & 2.490 & 0.013  & 5.229 & 0.022  & 1.844 & 0.031\\
V24 & 0.478471 & 0.85 & 15.100 & 0.315 & 0.498 & 0.018  & 2.507 & 0.186  & 5.143 & 0.283  & 1.748 & 0.382\\
V27 & 0.470532 & 1.45 & 15.004 & 0.498 & 0.504 & 0.020  & 2.327 & 0.236  & 4.691 & 0.347  & 0.951 & 0.465\\
V28 & 0.543865 & 0.97 & 15.121 & 0.328 & 0.500 & 0.007  & 2.348 & 0.016  & 5.035 & 0.022  & 1.463 & 0.030\\
V29 & 0.451332 & 0.88 & 15.164 & 0.359 & 0.365 & 0.006  & 2.231 & 0.049  & 4.487 & 0.072 & -0.008 & 0.116\\
V30 & 0.592207 & 0.82 & 15.093 & 0.278 & 0.496 & 0.008  & 2.429 & 0.042  & 5.284 & 0.063  & 1.854 & 0.083\\
V32 & 0.457797 & 1.31 & 15.146 & 0.452 & 0.462 & 0.002  & 2.233 & 0.006  & 4.680 & 0.009  & 0.997 & 0.014\\
V33 & 0.501575 & 1.17 & 15.128 & 0.380 & 0.489 & 0.039  & 2.228 & 0.109  & 4.736 & 0.142  & 0.898 & 0.181\\
V34 & 0.568119 & 0.82 & 15.087 & 0.293 & 0.447 & 0.015  & 2.288 & 0.038  & 4.992 & 0.060  & 1.682 & 0.072\\
V38 & 0.470437 & 0.86 & 15.114 & 0.342 & 0.444 & 0.006  & 2.522 & 0.026  & 5.232 & 0.042  & 1.820 & 0.058\\
V39 & 0.589035 & 1.17 & 14.999 & 0.389 & 0.522 & 0.003  & 2.431 & 0.009  & 5.181 & 0.013  & 1.687 & 0.018\\
V41 & 0.488577 & 1.06 & 15.140 & 0.404 & 0.446 & 0.009  & 2.213 & 0.035  & 4.851 & 0.047  & 1.202 & 0.062\\
V43 & 0.660177 & 0.61 & 15.047 & 0.224 & 0.464 & 0.005  & 2.656 & 0.021  & 5.563 & 0.033  & 2.484 & 0.056\\
V45 & 0.616632 & 1.00 & 14.994 & 0.324 & 0.540 & 0.010  & 2.423 & 0.028  & 5.310 & 0.041  & 1.790 & 0.057\\
V47 & 0.539739 & 1.04 & 15.134 & 0.348 & 0.454 & 0.012  & 2.316 & 0.067  & 4.960 & 0.100  & 1.409 & 0.132\\
V52 & 0.501785 & 1.07 & 14.973 & 0.373 & 0.488 & 0.020  & 2.979 & 0.164  & 4.745 & 0.182  & 3.194 & 0.502\\
V54 & 0.454239 & 1.19 & 15.051 & 0.427 & 0.457 & 0.008  & 2.250 & 0.050  & 4.770 & 0.074  & 1.072 & 0.100\\
V56 & 0.534849 & 0.62 & 15.135 & 0.255 & 0.384 & 0.004  & 2.353 & 0.040  & 5.414 & 0.062  & 2.342 & 0.092\\
V59 & 0.542027 & 0.99 & 14.975 & 0.324 & 0.509 & 0.004  & 2.339 & 0.014  & 5.076 & 0.021  & 1.486 & 0.030\\
V61 & 0.568647 & 0.91 & 15.097 & 0.312 & 0.506 & 0.004  & 2.402 & 0.011  & 5.179 & 0.016  & 1.662 & 0.021\\
V63 & 0.497993 & 0.70 & 15.263 & 0.565 & 0.497 & 0.081  & 3.449 & 0.148  & 7.221 & 0.244  & 0.754 & 0.284\\
V64 & 0.544492 & 1.02 & 15.134 & 0.337 & 0.504 & 0.003  & 2.336 & 0.017  & 5.009 & 0.025  & 1.445 & 0.037\\
V65 & 0.480758 & 0.96 & 15.117 & 0.363 & 0.446 & 0.012  & 2.409 & 0.047  & 5.004 & 0.067  & 1.219 & 0.098\\
V74 & 0.453987 & 1.39 & 15.080 & 0.464 & 0.489 & 0.028  & 2.233 & 0.093  & 4.657 & 0.136  & 0.859 & 0.199\\
V75 & 0.685471 & 0.55 & 15.002 & 0.208 & 0.442 & 0.010  & 2.698 & 0.041  & 5.635 & 0.065  & 2.708 & 0.123\\
V77 & 0.845261 & 0.60 & 14.773 & 0.234 & 0.415 & 0.005  & 2.934 & 0.021  & 6.238 & 0.044  & 3.539 & 0.123\\
V81 & 0.557292 & 0.94 & 15.087 & 0.331 & 0.456 & 0.024  & 2.420 & 0.061  & 4.999 & 0.089  & 1.526 & 0.052\\
V82 & 0.558927 & 0.92 & 15.035 & 0.298 & 0.544 & 0.008  & 2.446 & 0.046  & 5.192 & 0.065  & 1.710 & 0.088\\
V87 & 0.739210 & 0.35 & 14.922 & 0.150 & 0.353 & 0.007  & 2.850 & 0.036  & 6.021 & 0.064  & 3.365 & 0.118\\
V89 & 0.558454 & 0.94 & 15.114 & 0.316 & 0.481 & 0.010  & 2.224 & 0.038  & 4.937 & 0.054  & 1.481 & 0.080\\
V91 & 0.601589 & 1.31 & 14.850 & 0.361 & 0.640 & 0.111  & 2.367 & 0.255  & 5.400 & 0.402  & 1.344 & 0.490\\
V963 & 0.766991 & 0.20 & 14.949 & 0.093 & 0.194 & 0.011  & 2.609 & 0.104 & 7.254 & 0.255  & 3.160 & 0.481\\
\hline
Mean & 0.554059 & 0.97 & 15.066 & 0.346 & 0.470 & -      & 2.429 & -     & 5.172 & - & 1.634 & -\\
\hline
\hline
\end{tabular}
\end{flushleft}
\end{table*}

\begin{table}
\caption[ ]{Physical parameters of the RRab variables in M5}
\begin{flushleft}
\begin{tabular}{|r|c|c|c|c|c|r|c|}
\hline
\hline
Star & $M_V$ & $\sigma_{M_V}$ & [Fe/H] & $\sigma_{\rm [Fe/H]}$ &
$T_{eff}$ & $D_m$ \\
\hline
\hline
V1   & 0.841  & 0.085 & -1.092  & 0.068 & 6542 &  1.38\\
V2   & 0.824  & 0.086 & -1.087  & 0.168 & 6553 &  3.35\\
V3   & 0.814  & 0.093 & -1.133  & 0.045 & 6377 &  1.09\\
V4   & 0.903  & 0.081 & -1.026  & 0.206 & 6626 &  2.67\\
V5   & 0.806  & 0.085 & -1.289  & 0.089 & 6461 &  3.31\\
V6   & 0.870  & 0.094 & -0.772  & 0.215 & 6539 &  4.13\\
V8   & 0.820  & 0.086 & -1.210  & 0.049 & 6476 &  0.92\\
V9   & 0.700  & 0.101 & -1.174  & 0.076 & 6298 &  2.49\\
V10  & 0.821  & 0.084 & -1.241  & 0.078 & 6506 &  0.79\\
V11  & 0.747  & 0.089 & -1.274  & 0.026 & 6429 &  2.31\\
V12  & 0.854  & 0.078 & -1.159  & 0.049 & 6614 & 13.93\\
V13  & 0.814  & 0.080 & -1.483  & 0.100 & 6481 &  1.84\\
V14  & 0.867  & 0.082 & -1.309  & 0.221 & 6533 &  7.47\\
V16  & 0.670  & 0.091 & -1.497  & 0.037 & 6332 &  4.75\\
V18  & 0.813  & 0.074 & -1.452  & 0.061 & 6604 &  2.96\\
V19  & 0.824  & 0.077 & -1.234  & 0.086 & 6628 &  2.77\\
V20  & 0.764  & 0.092 & -1.261  & 0.071 & 6372 &  2.44\\
V21  & 0.763  & 0.091 & -1.267  & 0.036 & 6393 &  1.04\\
V24  & 0.938  & 0.092 & -0.701  & 0.381 & 6600 &  4.27\\
V27  & 0.814  & 0.084 & -1.267  & 0.468 & 6634 &  3.67\\
V28  & 0.829  & 0.086 & -1.200  & 0.034 & 6480 &  1.32\\
V29  & 0.886  & 0.076 & -1.438  & 0.105 & 6638 &  6.83\\
V30  & 0.811  & 0.092 & -1.126  & 0.087 & 6409 &  0.75\\
V32  & 0.853  & 0.077 & -1.213  & 0.033 & 6628 &  1.11\\
V33  & 0.832  & 0.081 & -1.373  & 0.192 & 6542 &  1.69\\
V34  & 0.807  & 0.087 & -1.388  & 0.083 & 6384 &  2.28\\
V38  & 0.945  & 0.088 & -0.538  & 0.060 & 6650 &  3.21\\
V39  & 0.752  & 0.089 & -1.247  & 0.026 & 6449 &  2.65\\
V41  & 0.851  & 0.081 & -1.148  & 0.067 & 6583 &  1.71\\
V43  & 0.771  & 0.099 & -1.116  & 0.055 & 6289 &  1.24\\
V45  & 0.758  & 0.093 & -1.223  & 0.060 & 6401 &  1.62\\
V47  & 0.817  & 0.085 & -1.279  & 0.135 & 6477 & 14.53\\
V52  & 0.836  & 0.082 & -1.362  & 0.247 & 6307 & 35.10\\
V54  & 0.879  & 0.079 & -1.073  & 0.102 & 6645 &  1.46\\
V56  & 0.916  & 0.093 & -0.642  & 0.087 & 6495 & 13.62\\
V59  & 0.838  & 0.087 & -1.135  & 0.033 & 6492 & 14.87\\
V61  & 0.817  & 0.089 & -1.139  & 0.028 & 6453 &  0.71\\
V64  & 0.821  & 0.086 & -1.238  & 0.039 & 6476 & 14.83\\
V65  & 0.897  & 0.084 & -0.900  & 0.091 & 6628 &  3.96\\
V74  & 0.850  & 0.077 & -1.223  & 0.186 & 6647 &  4.21\\
V75  & 0.751  & 0.101 & -1.156  & 0.096 & 6237 &  2.16\\
V77  & 0.578  & 0.115 & -1.208  & 0.095 & 6075 &  4.94\\
V81  & 0.805  & 0.087 & -1.320  & 0.122 & 6439 &  2.87\\
V82  & 0.839  & 0.090 & -1.069  & 0.089 & 6464 &  3.52\\
V87  & 0.744  & 0.109 & -0.927  & 0.104 & 6173 &  3.64\\
V89  & 0.804  & 0.086 & -1.410  & 0.076 & 6414 &  2.25\\
V91  & 0.771  & 0.103 & -1.019  & 0.541 & 6520 &  8.83\\
V963  & 0.860  & 0.133  & 0.581  & 0.366 & 6510 &24.71\\
\hline
\hline
\end{tabular}
\end{flushleft}
\end{table}

\begin{table*}
\caption[ ]{Mean parameters for cluster RRab stars based on the  method
of KJ} 
\begin{flushleft}
\begin{tabular}{|l|c|c|c|c|c|c|c|}
\hline
\hline
Cluster & No. of & mean & mean &   mean     & mean   & mean     & Reference \\
        & stars  & $P$  & $\phi_{31}$ & $M_V$ & [Fe/H] & $T_{eff}$ &    \\
\hline
\hline
NGC 6171 & 3 & 0.536 & $5.22\pm0.18$ & $0.85\pm0.02$ & $-0.91\pm0.13$ & $6619\pm64$ & Clement and Shelton (1997) \\
M5      & 26 & 0.555 & $5.05\pm0.06$ & $0.81\pm0.01$ & $-1.23\pm0.03$ & $6465\pm22$ & this work \\
NGC 1851 & 7 & 0.555 & $5.10\pm0.06$ & $0.80\pm0.01$ & $-1.17\pm0.03$ & $6494\pm35$ & Walker (1999) \\
M3      & 17 & 0.562 & $4.95\pm0.05$ & $0.78\pm0.01$ & $-1.42\pm0.03$ & $6438\pm18$ & Kaluzny et al. (1998)\\
M55     &  3 & 0.655 & $5.21\pm0.22$ & $0.68\pm0.02$ & $-1.56\pm0.16$ & $6325\pm69$ & Olech et al. (1999a) \\ 
\hline
\hline
\end{tabular}
\end{flushleft}
\end{table*}

\section{Conclusions}

We have presented accurate $V$-band CCD photometry of 65 RR Lyr
variables in the globular cluster M5. Among these stars we have
detected 49 fundamental mode pulsators, 15 1st overtone pulsators and
one possible 2nd overtone pulsator.

Four of our RRc stars with shortest periods show remarkably similar
light curves with a clear bump visible just before the maxima and
increased scatter of the observational points around maximum light. We
suggest that these features, due to the short pulsation periods of the
stars, are caused by interaction between the first and the second
overtones of radial pulsations.

Our precise photometry allowed us to obtain accurate Fourier
coefficients for 14 RRc stars (twice as many as in previous
investigations) and to estimate the physical parameters of these
objects. We have measured the mean mass of these stars to be
0.54~${\cal M}_\odot$, the mean logarithm of luminosity $L/L_{\odot}$
equal to 1.69, the mean effective  temperature of 7353~K and the mean
relative helium abundance equal to 0.28. These values are consistent
with the previous determinations of physical parameters of M5 and also
with recent theoretical models (Reid 1996, Clement and Shelton
1997, Sandquist et al. 1996, Caputo et al. 1999).

As many as 26 out of the 49 observed  RRab stars have regular light
curves as measured by the $D_m$ parameter of KJ and thus are suitable
for estimating the absolute magnitude, metallicity and effective
temperature from the Fourier coefficients of these stars.  From this
sample we obtained $M_V=0.81$~mag, ${\rm [Fe/H]}=-1.23$ and
$T_{eff}=6465$~K. The derived metallicity corresponds to the Jurcsik
(1995) scale which was used by KJ for calibrating their method.
Jurcsik (1995) gave ${\rm [Fe/H]}=-1.25$ for M5 on her scale and this
value is in excellent agreement with our estimate.

We also found that six of the regular RRab variables  are in a more
advanced evolutionary state than the rest of our sample. A detailed
inspection of the $A_V - \log P$ relations for the RRab stars from M3
and M5 suggests that the $V$ amplitude of RRab variables at a given
period is a function of the metallicity rather than the Oosterhoff type
of the cluster.

We estimate distance moduli of $(m -M)_V$ = 14.47 for the RRc stars in our
sample, and $(m -M)_V$ = 14.27 for the RRab stars. A possible 
correlation between measured luminosity as derived from the Fourier
coefficients and apparent brightness was identified for both the RRc
and RRab stars in our sample.

\begin{acknowledgements} 
JK, AO and WK were supported by the Polish Committee of Scientific
Research through grant 2P03D-011-12 and by NSF grant AST-9528096 to
Bohdan Paczy\'nski. ASC acknowledges support by KBN grant 2P03C-001-12 and
WP was supported by KBN grant 2P03-010-15.
\end{acknowledgements}

\clearpage

\noindent {\bf Fig. 1} $V$-band light curves of RR Lyr variables in M5.
The stars are plotted according to the increasing period.

\includegraphics{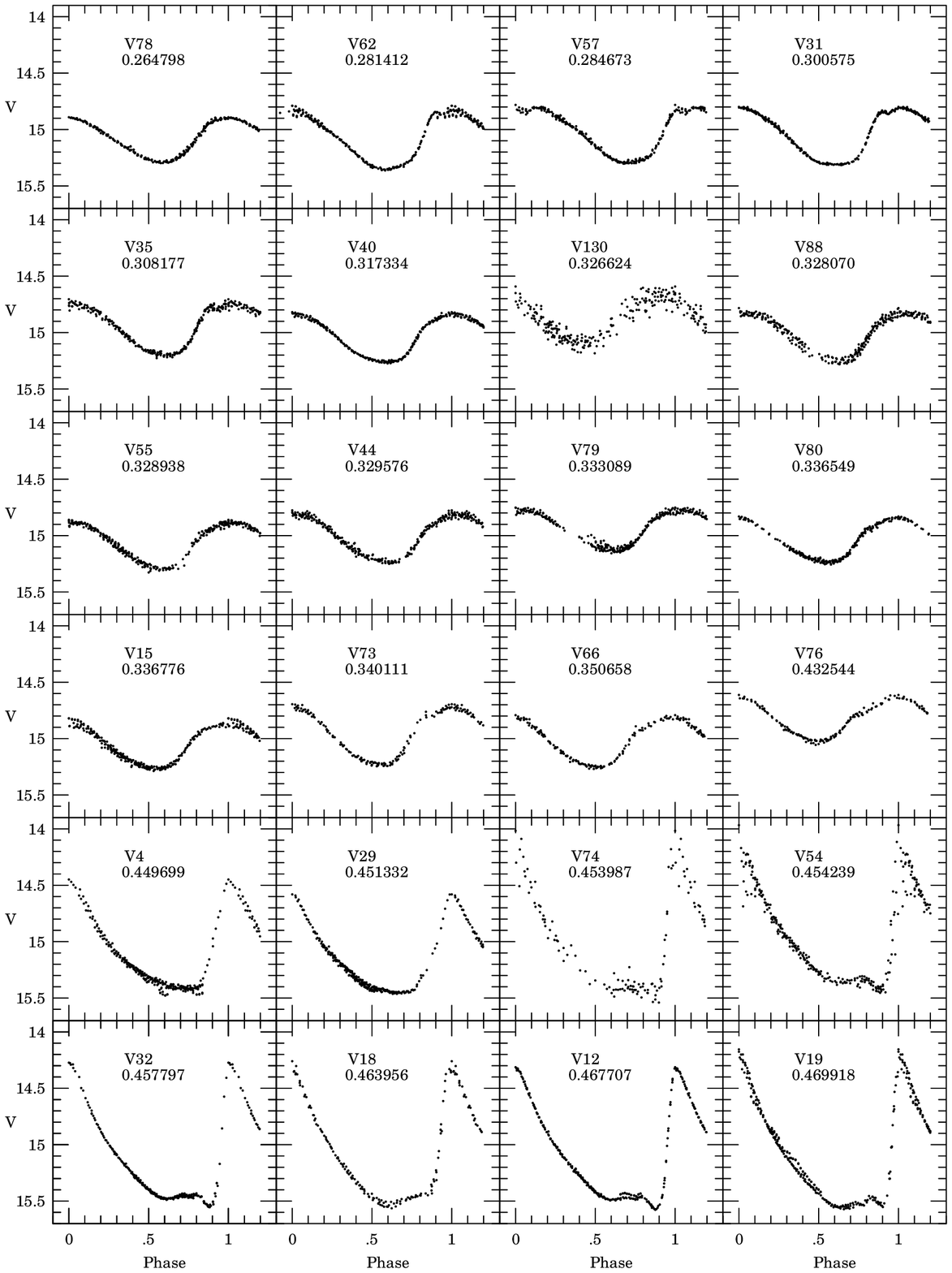}

\clearpage
 
\noindent {\bf Fig. 1} Continued
 
\includegraphics{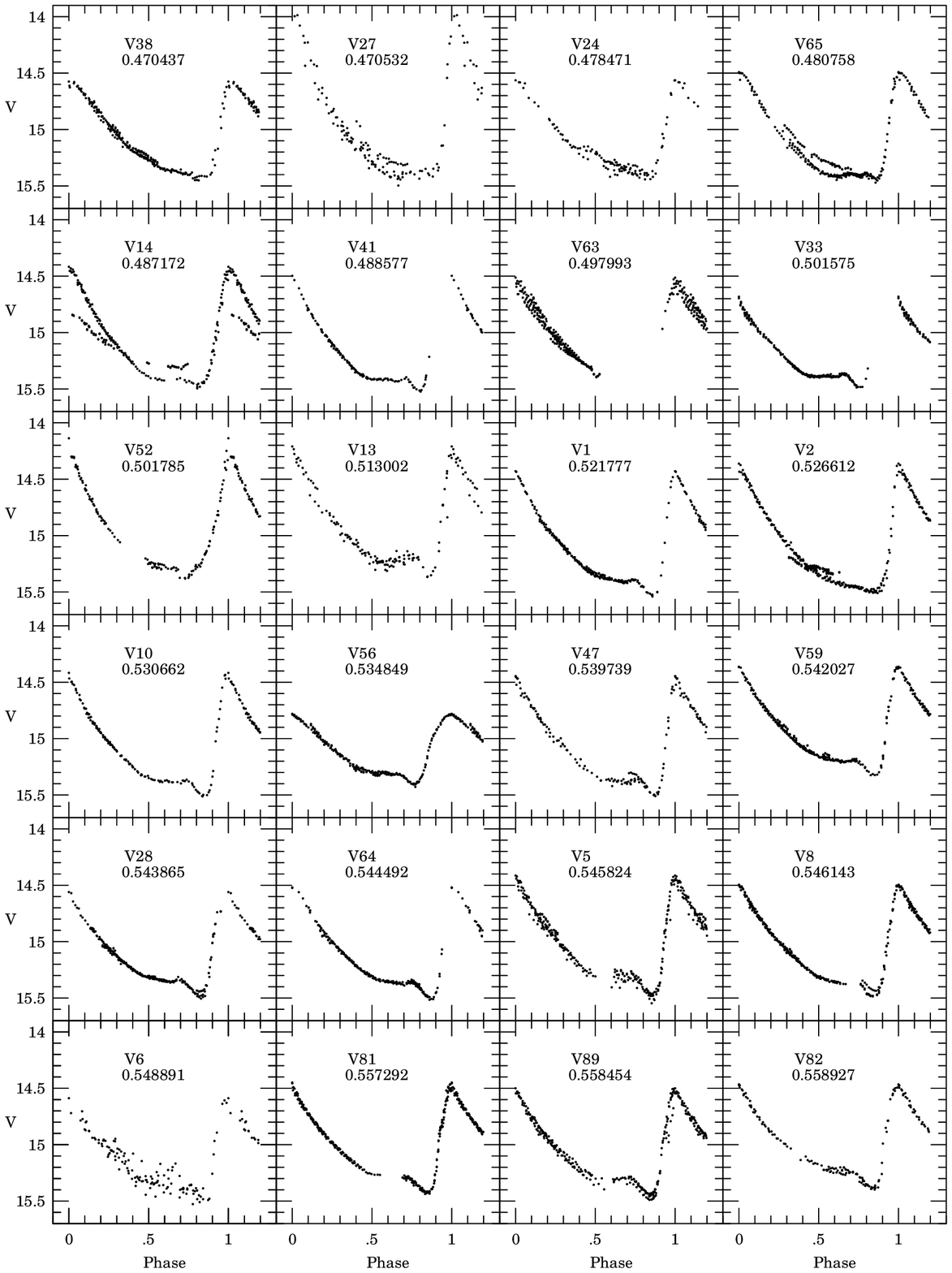}
 
\clearpage
 
\noindent {\bf Fig. 1} Continued
 
\includegraphics{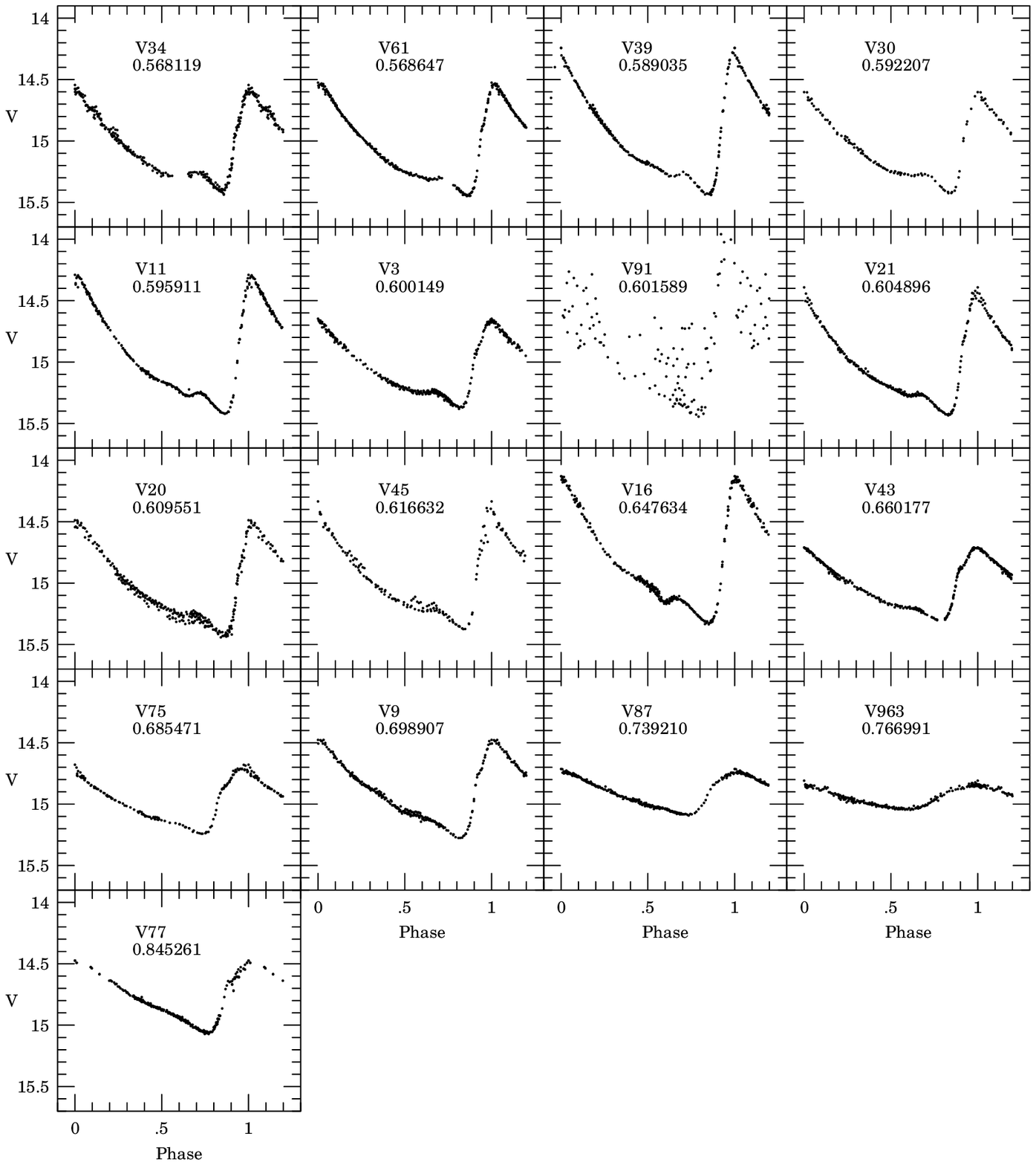}

\clearpage
 
\noindent {\bf Fig. 2} A comparison of the photometry presented in this
work with the observations by Reid (1996). RRc variables are plotted
with circles and RRab stars with triangles.
 
\includegraphics{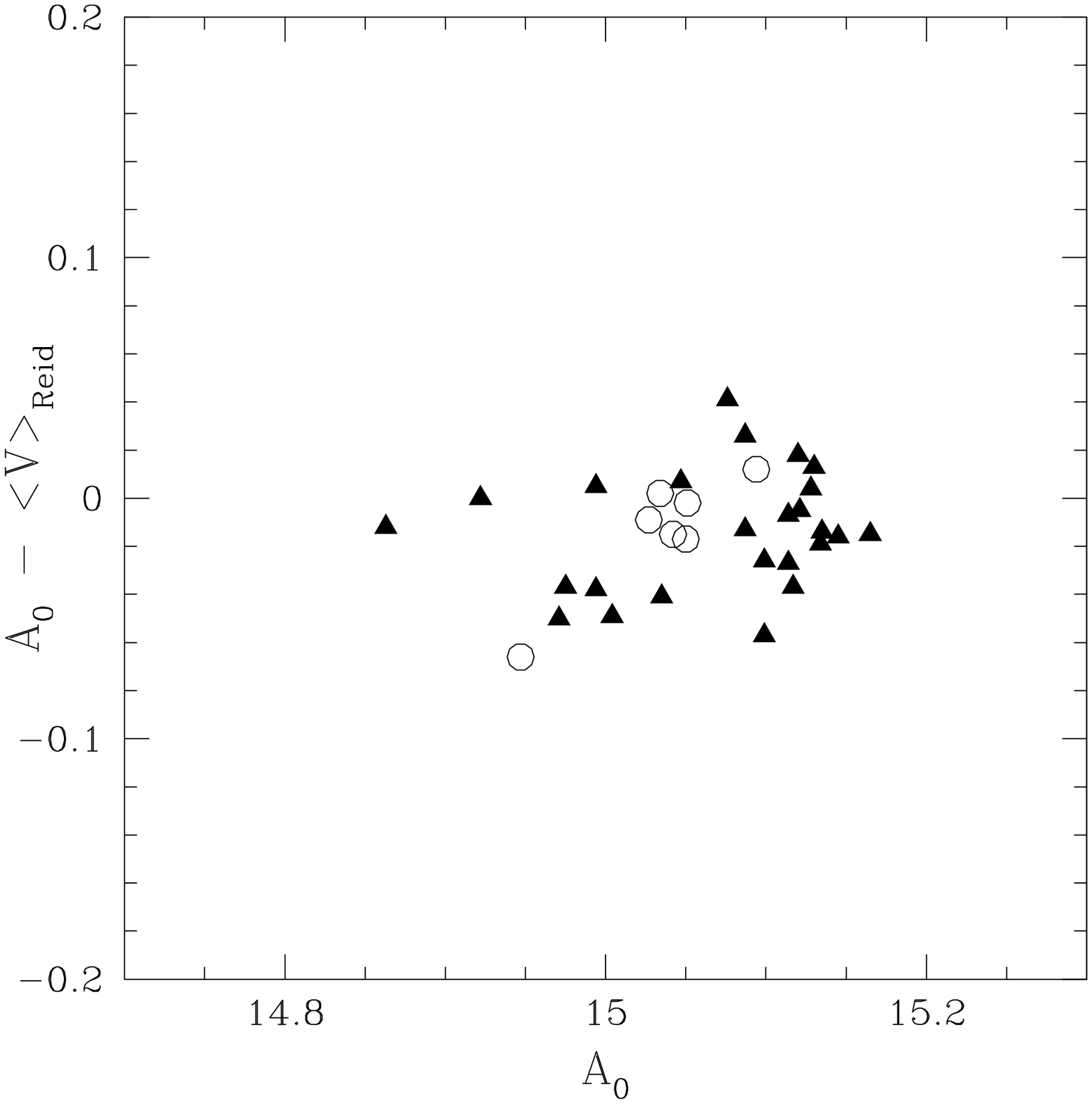}
 
\clearpage
 
\noindent {\bf Fig. 3} Amplitude, amplitude ratio and Fourier phase
differences as a function of period. RRc variables are plotted with
circles, RRab stars with $D_m<3$ with solid triangles and RRab variables
with $D_m>3$ with open triangles. The solid line in the upper panel of the
figure represents a linear fit to RRab variables in M3 (Kaluzny et al.
1998).
 
\includegraphics{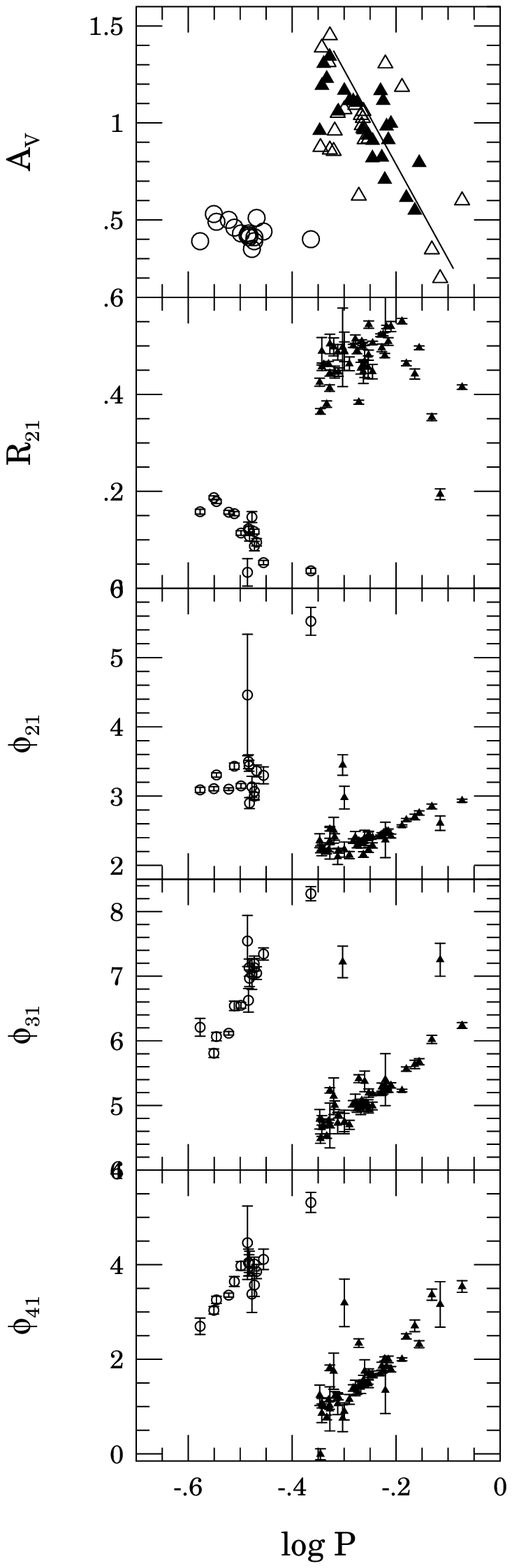}

\clearpage
 
\noindent {\bf Fig. 4} The dependence between luminosity and visual
magnitude for the RRc stars in M5. The solid lines have a slope of 0.4
and are separated by 0.04 in $\log L$, which represents the uncertainty
in the values of $\log L$ computed from $\phi_{31}$ and $P_1$. The size
of the circles corresponds to the distance from the center of the
cluster with larger circles laying at larger distances.
 
\includegraphics{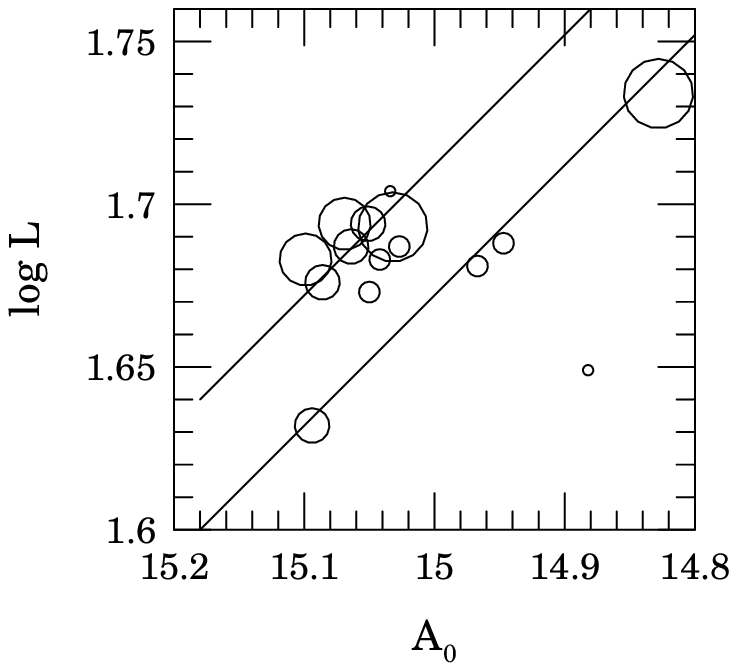}

\vspace{8.7cm}

\noindent {\bf Fig. 5} $M_V$ versus mean visual magnitude for the M5
RRab variables. The envelope lines, plotted with the slope of unity, are
separated by 0.1 mag in $M_V$, representing the uncertainty in the
derived magnitudes (Kovacs and Jurcsik 1996).

\includegraphics{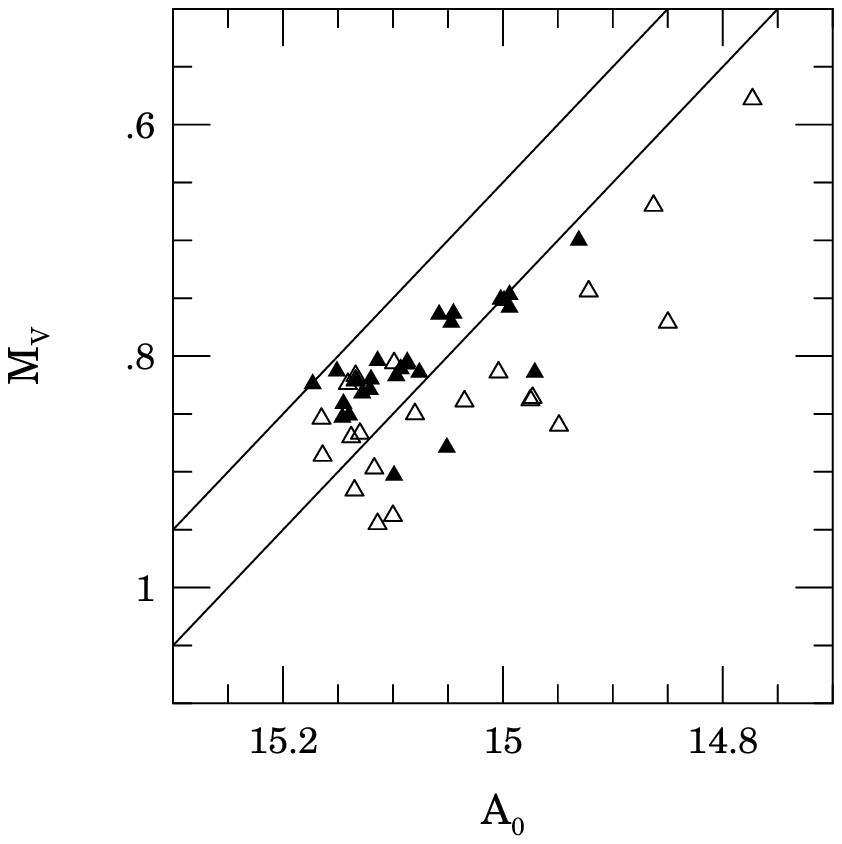}

\clearpage

\noindent {\bf Fig. 6} Apparent distance modulus plotted against $A_0$
for RRc stars (open circles), RRab stars with $D_m < 3.0$ (closed
triangles), and RRab stars with $D_m > 3.0$ (open triangles).

\includegraphics{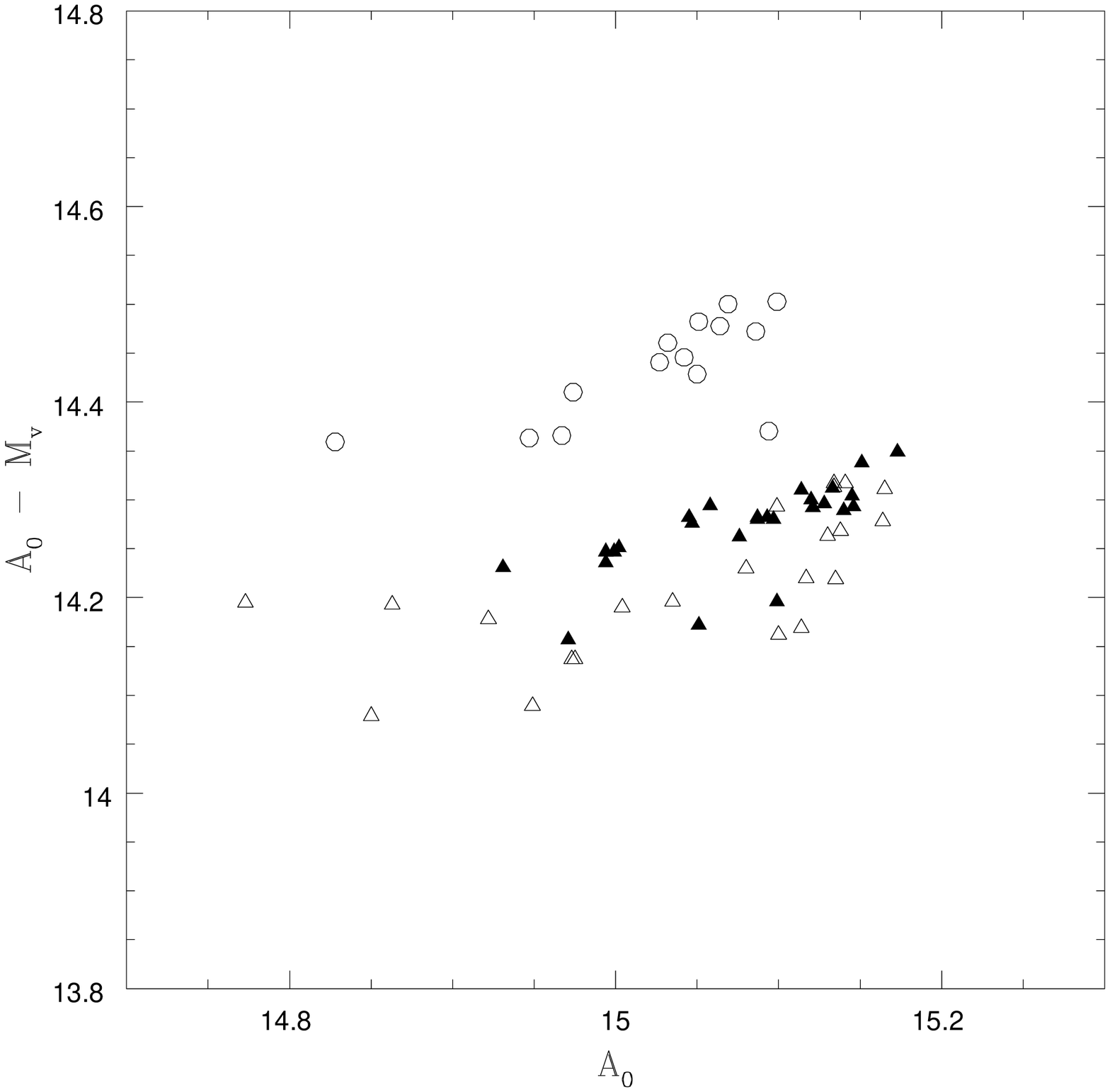}

\end{document}